\RequirePackage[2020-02-02]{latexrelease}
\documentclass[aps,floatfix,superscriptaddress,showpacs,twocolumn]{revtex4}

\usepackage{epsfig}
\usepackage{amsmath}
\usepackage{graphicx,psfrag}
\usepackage{dcolumn}
\usepackage{bm}
\usepackage{slashed}
\usepackage{color}

\newcommand{\beq}{\begin{equation}}
\newcommand{\eeq}{\end{equation}}
\newcommand{\bea}{\begin{eqnarray}}
\newcommand{\eea}{\end{eqnarray}}

\newcommand{\nn}{\nonumber\\}

\newcommand\fig[1]     {Fig.\,{\ref{#1}}}

\def\eq#1{(\ref{#1})}
\def\s0#1#2{\mbox{\small{$ \frac{#1}{#2} $}}}
\def\0#1#2{\frac{#1}{#2}}

\def\mr#1{{\mathrm{#1}}}

\begin{document}

\title{Polarised superlocalization in magnetic nanoparticle hyperthermia} 

\author{Zs. Isz\'aly} 
\affiliation{University of Debrecen, H-4010 Debrecen P.O.Box 105, Hungary}

\author{I. Gresits} 
\affiliation{Department of Biophysics and Radiation Biology, Semmelweis University, Budapest Hungary}
\affiliation{Institute of Physics, Budapest University of Technology and Economics, Po. Box 91, H-1521 Budapest, Hungary}

\author{I. G. M\'ari\'an}
\affiliation{University of Debrecen, H-4010 Debrecen P.O.Box 105, Hungary}
\affiliation{MTA-DE Particle Physics Research Group, P.O.Box 51, H-4001 Debrecen, Hungary}

\author{Gy. Thur\'oczy} 
\affiliation{Department of Non-Ionizing Radiation, National Public Health Center, Budapest}

\author{O. S\'agi}
\affiliation{Institute of Physics, Budapest University of Technology and Economics, Po. Box 91, H-1521 Budapest, Hungary}

\author{B. G. M\'arkus}
\affiliation{Institute of Physics, Budapest University of Technology and Economics, Po. Box 91, H-1521 Budapest, Hungary}
\affiliation{Stavropoulos Center for Complex Quantum Matter, Department of Physics, University of Notre Dame, Notre Dame, Indiana 46556, USA}
\affiliation{Institute for Solid State Physics and Optics, Wigner Research Centre for Physics, PO. Box 49, H-1525, Hungary}

\author{F. Simon} 
\affiliation{Institute of Physics, Budapest University of Technology and Economics, Po. Box 91, H-1521 Budapest, Hungary} 
\affiliation{Institute for Solid State Physics and Optics, Wigner Research Centre for Physics, PO. Box 49, H-1525, Hungary}

\author{I. N\'andori}
\affiliation{University of Debrecen, H-4010 Debrecen P.O.Box 105, Hungary}
\affiliation{MTA-DE Particle Physics Research Group, P.O.Box 51, H-4001 Debrecen, Hungary}
\affiliation{Atomki, P.O. Box 51, H-4001 Debrecen, Hungary} 

\begin{abstract} 
Magnetic hyperthermia is an adjuvant therapy for cancer where injected magnetic nanoparticles 
are used to transfer energy from the time-dependent applied magnetic field into the surrounding 
medium. Its main importance is to be able to increase the temperature of the human body locally.
This localization can be further increased by using a combination of static and alternating external 
magnetic fields. For example, if the static field is inhomogeneous and the alternating field is
oscillating then the energy transfer and consequently, the heat generation is non-vanishing only
where the gradient field is zero which results in superlocalization. Our goal here is to 
study theoretically and experimentally whether the perpendicular or parallel combination of static 
and oscillating fields produce a better superlocalization. A 
considerable polarisation effect in superlocalization for small frequencies and large field strengths
which are of great importance in practice is found.
\end{abstract}

\maketitle

\section{Introduction}
\label{sec_intro}
In cancer therapy local, heat generation is of great importance since tumor cells are more 
sensible for temperature increase than normal ones. Magnetic nanoparticles (MNPs) injected 
into the human body can transfer energy from the external applied time-dependent magnetic 
field into their environment thus can be used to increase the temperature locally 
\cite{pankhurst,ortega_d,pankhurst_progress,perigo_e_a,cabrera_d,clinapp_1,clinapp_2}.
A recent review can be accessed under Ref.~\cite{review_mnp}.
However, these MNPs can spread not just in cancerous tissues 
but can reach normal healthy cells. 
Therefore, there is a need to localize further the heating efficiency which can be done by using 
a combination of a gradient static and a time-dependent alternating magnetic fields since 
for large enough static field the dissipation is expected to be dropped to zero. As a consequence, 
the temperature increase is observed only where the static field vanishes \cite{mpi_test}. There is an 
increasing interest in the literature see e.g.,~\cite{adv_func_mat,review_mnp} on how to 
improve the efficiency of the method, and how to "superlocalize" the heat transfer 
\cite{non_local_heat,focused_hyperthermia,recent_focused_hyperthermia,stat_along_increased}. 
The idea of superlocalization is based on that if the static field is present, the amount 
of energy transferred is decreased and one expects a bell-shaped curve of the energy transfer 
as a function of the static field amplitude, see e.g. Fig.~(10.4) of \cite{review_mnp} or 
for more details \cite{MRSh_theory_focused}.

The superlocalization effect is studied recently in \cite{superlocal_thermal} employing theoretical 
methods in particular by the deterministic  \cite{LLG} and stochastic \cite{sLLG}
Landau-Lifshitz-Gilbert (LLG) equations. It was shown that the use of the stochastic solver is 
important if one would like to compare the efficiency and the superlocalization of various
types of applied fields. For example, when the orientation of the static and alternating fields
and in addition the type (oscillating or rotating) of the time-dependent magnetic field are 
chosen differently. Results of Ref.~\cite{superlocal_thermal} suggest that the polarisation of the 
static field, which can be parallel or perpendicular to the direction of oscillation matters. 
However, it is not clear how this polarisation effect depends on the parameters, i.e., 
the applied frequency, damping parameter, etc. 

In this work, our goal is to study the details of the polarisation effect, i.e., whether the 
parallel or the perpendicular combination of static and oscillating fields give a better
heating efficiency and a better superlocalization. The parameter dependence of the method 
(applied frequency, damping parameter), the validity of approximations is examined herein. 
The results are then compared with the experimental results performed.
We demonstrate that a considerable polarisation effect in superlocalization is found, which 
is of great importance in practice.

This paper is centered around the mentioned questions and is organized as follows.  
After the introduction (Sec.~\ref{sec_intro}) we discuss the theoretical framework
in Sec.~\ref{sec_th}, and present the theoretical results where the focus is on the
difference of the energy losses obtained for the parallel and perpendicular combination
of the static and the oscillating fields. As a next step, in Sec.~\ref{sec_exp}, we explain
the details of the experimental setup used to investigate the polarization effect of the cases 
mentioned above, where the static and the time-dependent fields
are chosen to be either parallel or perpendicular. In Sec.~\ref{sec_compare}, we 
compare the theoretical predictions and experimental data. Finally,  Sec.~\ref{sec_sum} 
stands for the summary.

\section{Theoretical study of superlocalization}
\label{sec_th}
To study the efficiency and the superlocalization effect of magnetic hyperthermia, one 
should choose a theoretical framework and appropriate approximations. MNPs may
form clusters but typically the aggregation is not favored and can be neglected 
\cite{usov_claster,review_mnp}. Therefore, the investigation of single isolated MNPs is satisfactory. 

Another issue is the mechanical motion (rotation) of the particles in their environment. 
If the applied frequency is high enough ($f \gtrsim 100$ kHz) and the diameter 
of the nanoparticle is relatively small ($\sim 20$ nm),  the mechanical rotation of the MNPs 
can be restricted in the surrounding medium, and only the orientation of their magnetic 
moment has to be taken into account  \cite{usov_hysteresis,lyutyy_general,neel_brown,viscous_rotating}. 
Important to note that the frequency cannot be chosen to be too high to minimize eddy currents 
and to enable the use of the method for hyperthermic treatments. For example, a typical 
maximum value is around  $f \sim 500 - 1000$ kHz. Thus, if the applied frequency is 
chosen to be around $500$ kHz, the energy loss depends on the dynamics of the magnetization 
only and this can be described by the so-called deterministic \cite{LLG}  and stochastic \cite{sLLG} 
Landau-Lifshitz-Gilbert (LLG) equation. 

As a consequence, the frequency and the amplitude of the applied field, whose product is 
proportional to the power injected, has an upper bound, the Hergt-Dutz limit \cite{hergt_dutz}. 
General advice to exploit the heating potential 
of the particles is to use a field frequency of several hundred kHz in combination with a rather 
low field amplitude (few kA/m) for superparamagnetic particles and a relatively high field 
amplitude (a few tens of kA/m) in combination with a frequency of a few hundred kHz for MNPs 
with hysteretic behavior. Since the heat transfer depends on both the frequency and the 
amplitude almost linearly, it is a good choice to keep them at their maximum values.

Moreover, the diameter of the MNP matters \cite{usov_claster}, because as the volume 
is present in the stochastic description but we do not investigate the dependence of our
results on the diameter of the individual particles. Realistically, it cannot be too small since 
then the MNPs will not be stored in the human body for long enough, but it also cannot be 
too large, because then particles would have multiple magnetic domains, which is not in the 
favor regarding the heating efficiency. In general, a good choice for the diameter is 
between 10 and 50 nm, and we adopt this value in our analysis \cite{usov_claster}. 

A further approximation is related to the shape and crystal anisotropy of the MNPs. 
In principle, the inclusion of the anisotropy is straightforward in the theoretical approach 
but its realization in practice is certainly more difficult. It is very questionable 
whether the orientation of an oblate or prolate MNP in the human body can be supported
by any experimental realization. Instead, spherically symmetric particles require
no special attention and technics. Thus, here we restrict the theoretical study for the 
isotropic case where no shape and no crystal anisotropy are considered.

Taking into account the above mentioned approximations, we use the stochastic LLG-equation 
to study the motion of the magnetization vector of a single, isotropic MNP
which undergoes the so-called N\'eel relaxation. 
The thermally induced and/or forced magnetic dynamics of magnetic nanoparticles 
are being studied for a long time where the interest is stimulated by experimental 
evidence \cite{thermal_exp}. An excellent overview of the stochastic 
dynamics can be found in Ref.~\cite{thermal_summary}. 
The stochastic LLG equation \cite{sLLG} where thermal fluctuations are taken into 
account by introducing a random magnetic field, ${\bf H} = (H_x, H_y, H_z)$ reads as 
\begin{equation}
\label{sLLG}
\frac{\rm{d}}{{\rm{d}}t} {\bf M} = -\gamma' [{\bf M \times (H_{\rm{eff}}+H)}] 
+ \alpha' [[{\bf M\times (H_{\rm{eff}}+H)]\times M}].
\end{equation}
Here, the Cartesian components of the stochastic field are independent Gaussian white 
noise variables,
\begin{equation}
\label{stochastic_field}
\langle H_i(t) \rangle = 0, \hskip 0.5 cm \langle H_i(t_1) H_j(t_2) \rangle = 2 \, D \, \delta_{ij} \, \delta(t_1 -t_2)
\end{equation}
with $i = x, y, z$. Furthermore, $\gamma' = \mu_0 \gamma_0 /(1+\alpha^2)$, $\alpha' = \gamma' \alpha$ 
with the dimensionless damping $\alpha = \mu_0\gamma_0\eta$ with a damping factor 
$\eta$ and $\gamma_0 = 1.76 \times 10^{11}$~Am$^2$/Js  is the gyromagnetic ratio, while 
$\mu_0 = 4 \pi \times 10^{-7}$~Tm/A (or N/A$^2$) is the vacuum permeability. We introduced
the unit vector of the magnetic moment, ${\bf M} = {\bf m}/m_S$, where {\bf m} stands for the 
magnetization vector of a 
single-domain particle normalized by the saturation magnetic moment, $m_S$. 
For example, a typical value for the saturation magnetic moment is $m_S \approx 10^5$ A/m 
for a single crystal Fe$_3$O$_4$ \cite{Fannin}.  In addition, $D$ is 
a parameter which corresponds to the fluctuation-dissipation theorem, e.g. in Ref.~\cite{path_int_sllg}, 
that is defined as $D = \eta k_B T/(m_s V \mu_0)$ with the Boltzmann factor, $k_B$, the  absolute temperature, 
$T$, and the volume of the particle, $V$. The angular brackets stand for averaging over all possible 
realization of the stochastic field, ${\bf H}(t)$, and $\delta(t)$ is the Dirac $\delta$ function. 

Setting the stochastic field ${\bf H}(t)$ to zero yields the deterministic LLG equation. In case of an 
oscillating effective magnetic field, or more generally when the external field depends on $t$ through 
a function of $\omega t$, then the deterministic LLG equation is invariant under the following simultaneous 
transformations
\beq
\omega \to k \omega, \hskip0.5cm t \to t/k, \hskip0.5cm H\to k H.
\eeq
Here, $k$ is a well-chosen normalization parameter. It is important to note, that the same is not true for 
the stochastic LLG equation, since the stochastic field breaks this symmetry. To preserve the form of the 
stochastic LLG equation under these transformations, 
the temperature should be rescaled as well $T\to k T$.

The effective magnetic field contains the applied field which is the combination of static and oscillating ones:
\bea
\label{H_osc_parallel_def}
\mr{Parallel \,  oscillating} \hskip -0.07cm :  {\bf H}_{\rm{eff}} = H \, \Big(\cos(\omega t) + b_0,\,\,  0, \,\, 0\Big), \hskip 0.4cm \\
\label{H_osc_perp_def}
\mr{Perpendicular \, oscillating} \hskip -0.07cm : {\bf H}_{\rm{eff}} = H \, \Big(\cos(\omega t),\,\,  b_0, \,\, 0\Big), \hskip 0.4cm
\eea
where $H$ and $H b_0$ stands for the amplitude of the applied and the static fields, respectively.

Let us now discuss the parameters of the stochastic LLG equation \eq{sLLG}. In particular, we are interested 
in parameters that can be used for medical applications, i.e., which are suitable for magnetic hyperthermia. 
As mentioned earlier, the product of the frequency and the amplitude of the applied field has to be chosen to 
its maximum since it is related to the power injected and we want to keep it as much high as possible with 
respect to the Hergt-Dutz limit. The upper bound of $5 \times 10^{8}$ A/(m s) is used as a safe operational 
guideline \cite{review_mnp},  however, depending on the seriousness of the illness and the 
diameter of the exposed body region this critical product may be exceeded implying a higher upper bound of 
$5 \times 10^{9}$ A/(m s) \cite{hergt_dutz}. Another, less strict upper limit can be found in clinical applications 
\cite{magforce}, where the maximum of the applied frequency is $f = 100$ kHz and the maximum 
field strength  is $H = 18$ kA/m. Moreover, in the experimental study of Bordelon \emph{et al}. 
\cite{high_amplitude}, relatively high amplitudes of $H= 100$ kA/m at $f = 150$ kHz 
were utilized. In this work, we choose the following upper limit:
$\omega= 2\pi f = 2000$ kHz (i.e., $f \approx$ 300 kHz) and $H = 18$ kA/m. 

In addition, it is useful to introduce the following frequency-like parameters, $\omega_L =  H \gamma'$ 
and $\alpha_N = H \alpha'$. The typical values for the damping parameter are $\alpha = 0.1$ 
and $\alpha = 0.3$, which are used in Ref.~\cite{Lyutyy_energy} and in Ref.~\cite{Giordano}, respectively.
Thus, if $\alpha=0.1$ is taken (and $H = 18$ kA/m), a good and typical choice for 
parameters suitable for hyperthermia is $\omega_L = 4 \times10^9 \, \mr{Hz}$ and 
$\alpha_N =  4 \times 10^8 \, \mr{Hz}$. The following dimensionless parameters can be introduced 
to simplify the description:
\bea
\label{dimless_param_def}
\omega &\to&  \omega t_0  \nn
\omega_L =   H \gamma' &\to& \omega_L t_0  \nn
\alpha_N = H \alpha' &\to& \alpha_N t_0,
\eea
where the dimension has been rescaled by a suitably chosen time parameter, $t_0$. 
Here we use $t_0 = 0.5 \times 10^{-10}$s. In this case, with $\alpha = 0.1$, the dimensionless parameters 
are  $\omega_L \sim  0.2$, and $\alpha_N \sim 0.02$. From now on, $t_0$ will be omitted and only 
$\omega_L$ and $\alpha_N$ will be noted.

If the solution ($\bf M_{\mr{sol}}$) of the LLG equation is given, then the energy loss in a single
cycle can be determined in the following way,
\beq
\label{def_loss}
E = \mu_0 m_S \int_{0}^{\frac{2\pi}{\omega}} {\rm{d}}t 
\left({\bf H}_{\rm{eff}} \cdot \frac{{\rm{d}}{\bf M_{\mr{sol}}}}{{\rm{d}}t} \right).
\eeq
Let us note, that Eq.~\eq{def_loss} contains the average magnetization, {\bf $\bf M$}, of 
a single particle. The energy given by Eq.~\eq{def_loss} is identical to the area of the dynamical 
hysteresis loop, and it is related to the imaginary part of the frequency-dependent susceptibility. 
We will use these relations over the discussions of the theoretical numerical results.
   
The product $E \cdot f$, where $E$ is the energy and $f$ the applied frequency, is related to the 
specific loss power (SLP) or specific absorption rate (SAR), which has a dimension of W/kg,
\beq
\label{sar}
E\cdot f \propto \mr{SAR} = \mr{SLP} = \frac{\Delta T \,\, c}{t} \, .
\eeq
Here, $\Delta T$ is the temperature increment, $c$ is the specific heat, and $t$ is the time of the 
heating period. From Eq.~\eq{sar} one can derive another useful quantity, the intrinsic loss power (ILP),
\beq
\label{ilp}
\mr{ILP} = \frac{\mr{SAR}}{H^2 f} \,  {\propto \frac{E}{H^2}} \, ,
\eeq
which are used to compare experimental and theoretical SAR values obtained for different 
values of $f$ and $H$. In the present work, our theoretical numerical results are usually given for the 
dimensionless value,
\beq
\label{EpH}
\frac{E}{2\pi\mu_0 m_s H}  \propto \mr{ILP} \, .
\eeq
This can be derived from Eqs.~\eq{def_loss} and \eq{ilp}, when the amplitude $H$ is kept constant. 
For varying $H$, the proportionality only holds if the lhs.~of Eq.~\eq{EpH} is divided by $H$ as in Eq.~\eq{ilp}.

Our strategy is based on the numerical solution of the stochastic LLG equation \eq{sLLG} 
which is independent of the initial conditions of the magnetization vector. Once this solution
is determined, the energy loss per cycle can be calculated by using Eq.~\eq{def_loss}. We plot this 
variation as a function of the magnitude of the static field, $b_0$. In this way, we can compare 
the efficiency and the superlocalization of the parallel \eq{H_osc_parallel_def} and 
the perpendicular \eq{H_osc_perp_def} cases for various parameters which are plotted
in \fig{fig3} and \fig{fig6}.

\subsection{Vanishing static field}

\begin{figure}[ht] 
\begin{center} 
\includegraphics[width=7.8cm]{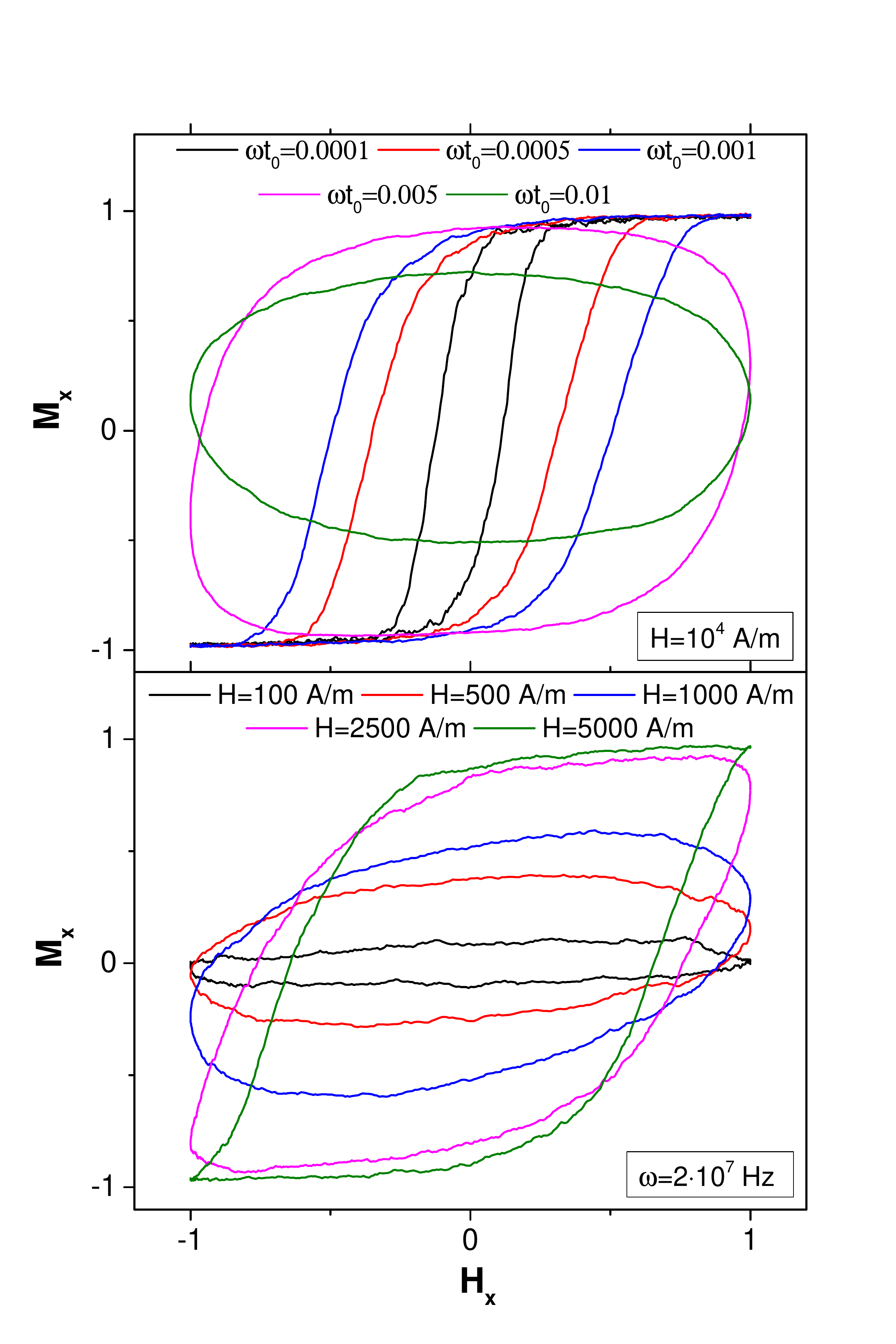}
\caption{(Color online.) Dynamical hysteresis loops are given (on the upper panel) for 
various angular frequencies of the applied AC magnetic field while the amplitude (field strength) 
is fixed, and (on the lower panel) for various field strengths of the applied AC magnetic 
field while the frequency is fixed. 
\label{fig1}
} 
\end{center}
\end{figure}

Let us first discuss the case of a vanishing static field. On the upper panel of \fig{fig1}, 
the dynamic hysteresis loops for various angular frequencies but at a fixed field strength is presented. 
%
%
First of all, the shape and the area of the loops depend on the frequency. 
It is also clear, from the upper panel of \fig{fig1}, that an optimal value for the 
frequency can be found, where the area of the hysteresis loop is the largest, i.e., one expects a peak 
of the energy loss as a function of the applied frequency (for fixed field strength).

Similar conclusions can be drawn from the lower panel of \fig{fig1}, where dynamic 
hysteresis loops are given for various field strengths, but for fixed frequency. 

Similarly to the previous figure, the shape and the area of the loops depend on the parameters 
of the applied magnetic field, in this case, the field strength. One finds parameters, where the 
shape of the loop is an ellipse (e.g., the blue-colored loop of the lower panel of \fig{fig1}), 
i.e., the averaged magnetization vector of a single nanoparticle has the following time 
dependence,
\beq
{\bf M} = M_0 \, \, \Big(\cos(\omega t - \varphi),\,\,  0, \,\, 0\Big).
\eeq
Thus, the computation of the imaginary, $\chi''$, and the real part, $\chi'$, of the magnetic AC 
susceptibility is straightforward. The imaginary part  of the susceptibility is related 
to the energy loss \eq{def_loss}, and is proportional to $\sin{\varphi}$, while its real part is related
to $\cos{\varphi}$. Furthermore, it is also shown that for certain field strength and angular frequency values, 
the shape of the loops differ from an ellipse caused by the saturation effect. Since both the shape 
and the area of the hysteresis loops are different, the corresponding AC susceptibility has an 
unusual dependence on the 
frequency and field strength. 

\begin{figure}[ht] 
\begin{center} 
\includegraphics[width=7.8cm]{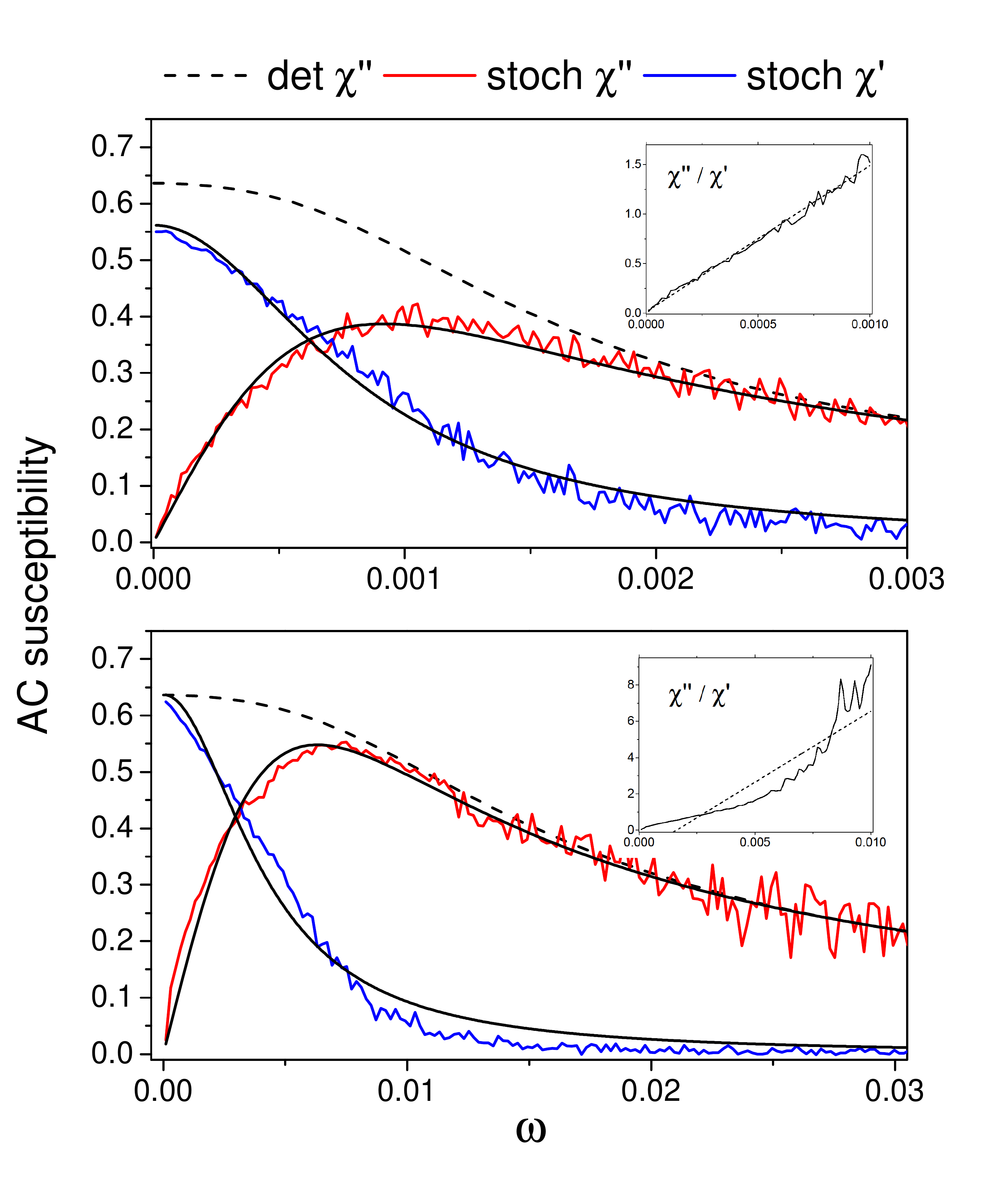}
\caption{(Color online.) The real and imaginary parts of the AC susceptibility as a function of 
the applied (dimensionless) angular frequency for $\alpha = 0.1$. The field amplitude is chosen 
to be identical to the curves presented in \fig{fig3}, i.e., on the upper panel $\alpha_N =0.002,\ \omega_L = 0.02$, 
and on the lower panel $\alpha_N =0.02,\ \omega_L = 0.2$. Dashed lines are the solution 
of the deterministic LLG equation. The solid lines are fits to numerical data based on the 
functions given by Eqs.~\eq{chi_real} and \eq{chi_imaginary}.
\label{fig2}
} 
\end{center}
\end{figure}

In \fig{fig2} the imaginary and the real parts of the AC susceptibility are presented as a function of 
the applied angular frequency (dimensionless). The field amplitude in the
upper and the lower panels are different, and it is chosen to be identical to the corresponding
panels of \fig{fig3}. Thus, the field strength is a magnitude larger on the lower panel of \fig{fig2},
compared to the value used in the upper panel.
%
%

An important observation is that the imaginary and real parts of the AC susceptibility of \fig{fig2} 
agree well with those presented in Fig.~2 of Ref.~\cite{ferguson}. 

If a single relaxation process is present, the real and imaginary parts of the complex magnetic 
susceptibility simplify reduces to the following forms (see e.g., Eq.~(4) and Eq.~(5) of Ref.~\cite{gresits3}),
\bea
\label{chi_real}
\chi'(\omega) &=& \chi_0 \frac{1}{1+\omega^2 \tau^2},  \\
\label{chi_imaginary}
\chi''(\omega) &=&  \chi_0 \frac{\omega \tau}{1+\omega^2 \tau^2}\ .
\eea
Here, the angular frequency dependence is thought to be described by the relaxation time of the 
nanoparticles, $1/\tau = 1/\tau_N + 1/\tau_B$, where the N\'eel and Brown relaxation times 
are related to the motion of the magnetization with respect to the particles and the motion of 
the particle itself, respectively. Since relatively small nanoparticles are considered, the motion 
of the particle as a whole can be neglected, thus only the N\'eel process counts and one can 
write $\tau \approx \tau_N$. Eqs.~\eq{chi_real} and \eq{chi_imaginary} are used to fit the numerical 
results obtained from the direct solution of the stochastic LLG equation. The solid lines of \fig{fig2} 
stand for these fitted curves. The deterministic LLG equation cannot reproduce the full curve 
described by Eq.~\eq{chi_imaginary}. However, for relatively large angular frequencies, ${\omega>1/\tau}$, 
it yields a quite accurate approximation, as noted by the dashed lines of \fig{fig2}.

The inset of the upper panel of \fig{fig2} shows that the ratio of the  real and imaginary 
parts obtained by the direct solution of the stochastic LLG equation is almost a straight line 
in the low-frequency regime. Therefore, we can conclude that the frequency-dependence 
of the susceptibility can be well described by Eqs.~\eq{chi_real} and \eq{chi_imaginary} if the applied 
field strength is small. In other words, for small field strength, the linear response theory can
be applied for the results of the stochastic LLG equation. In addition, our result gives 
$\tau_N \sim 70$ ns which is a typical value for the N\'eel relaxation time for magnetic 
nanoparticles ~\cite{ota}.

However, a different result can be obtained if the applied field is large. Indeed, the inset of the lower 
panel of \fig{fig2} demonstrates that the ratio of the real and imaginary parts obtained by the 
direct solution of the stochastic LLG equation, has considerable deviations from the straight 
line in the low-frequency range. Therefore, the linear response theory seems to be an adequate
approximation only for relatively large frequencies and small magnetic fields.

Accordingly, if the product of the frequency and the field strength is kept constant (to satisfy the 
Hergt-Dutz limit), the linear response theory cannot be applied for small frequencies (and for large 
field strengths) which is otherwise required for any medical treatments of magnetic hyperthermia.

\subsection{Finite static field, superlocalization}

Our main goal in this work is to consider whether the superlocalization effect depends 
on the particular choice of the parallel \eq{H_osc_parallel_def} and perpendicular 
\eq{H_osc_perp_def} combinations of static and oscillating fields. In particular, we 
study how sharp the superlocalization is in the parallel and perpendicular cases
for various parameters. 

As a first step, let us compare the superlocalization of the parallel and perpendicular 
cases at relatively high angular frequency ($2 \times 10^7$Hz) and small field strength 
($1.8$~kA/m).
%
%
\begin{figure}[ht] 
\begin{center} 
\includegraphics[width=7.8cm]{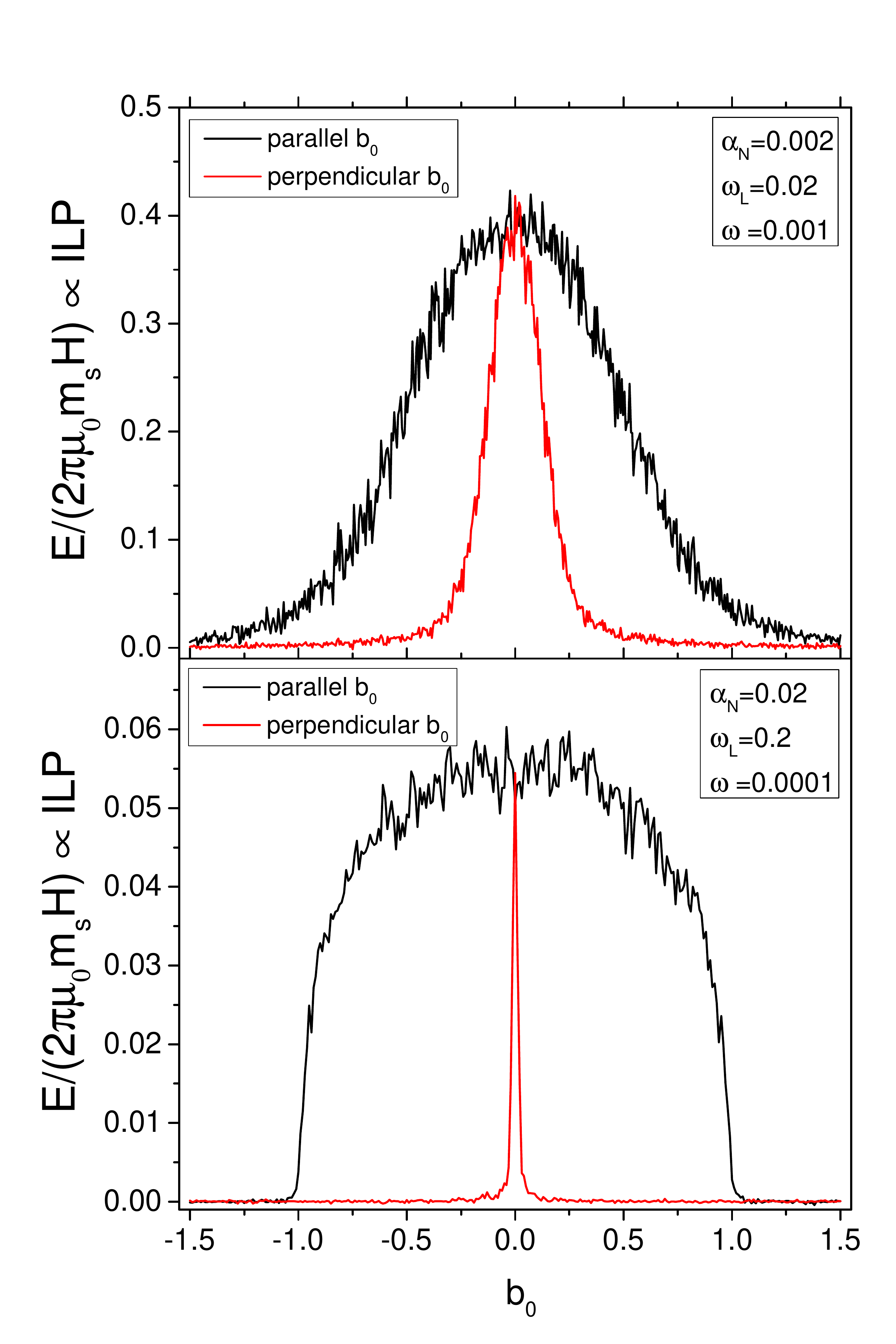}
\caption{(Color online.) On the upper panel, we compare the superlocalization 
effect of the parallel \eq{H_osc_parallel_def} and perpendicular combination 
\eq{H_osc_perp_def} of the static and oscillating fields averaged over 34
cycles. We choose the following dimensionful parameters, $\omega = 2 \times 10^7$ Hz, 
$H=1800$ A/m where the dimensionless damping is $\alpha = 0.1$. On the lower panel, 
one finds the same but the following dimensionful parameters suitable for hyperthermia, 
$\omega=2 \times 10^6$ Hz, $ H=18000$ A/m.
\label{fig3}} 
\end{center}
\end{figure}
In the upper panel of \fig{fig3}, the ILP of the parallel \eq{H_osc_parallel_def} 
and the perpendicular \eq{H_osc_perp_def} cases are plotted as a function of the static field 
amplitude or more precisely, $b_0$. The figure shows a moderate
polarisation effect, i.e., the superlocalization is better for the perpendicular case.

As a second step, we compare the superlocalization of the parallel and perpendicular 
cases at small angular frequency ($2 \times 10^6$Hz) and large field strength ($18$~kA/m), 
as depicted in the lower panel of \fig{fig3}. The product of the frequency and the field strength
is the same in the upper and lower panels of \fig{fig3}. An important observation is 
that the superlocalization effect is more pronounced in the lower panel of \fig{fig3} 
than in the other case. In both cases, the superlocaization is better for the 
perpendicular combination. 

Let us note, the presented theoretical results for the parallel case (see the black curves
of \fig{fig3}) agree well with previous literature, see for example, Figs.~10.3 and 10.4 of the review 
\cite{review_mnp}. Those figures are taken from Ref.~\cite{MRSh_theory_focused} where the 
results were obtained by solving the Martensyuk, Raikher, and Shliomis (MRSh) equation. 
Thus, Ref.~\cite{MRSh_theory_focused} represents a different theoretical framework which 
leads to the same superlocalization for the parallel case, i.e., if the AC and DC (bias) magnetic 
fields are parallel to each other, energy loss can only be observed if the magnitude of the DC field 
is smaller than or identical to the AC field.

Concludingly, a considerable polarisation effect in superlocalization of the parallel and 
perpendicular cases in the range of hyperthermia is observed, i.e., for small frequencies and large
field strengths. We showed in the previous subsection that this is the case where the 
linear response theory cannot be applied.

The theoretical results presented in \fig{fig3} suggest that for magnetic hyperthermia the 
superlocalization effect is much stronger if the static and the oscillating fields are perpendicular 
to each other. Notably, the perpendicular combination seems to be a better choice for magnetic 
hyperthermia as determined from calculations. As the conclusion of our theoretical considerations 
is obtained for a particular damping parameter, $\alpha = 0.1$ and based on various approximations, 
experimental verification is required, which is discussed in the next section. Furthermore, it has to be 
shown whether our conclusion remains unchanged if $\alpha$ is increased or decreased (of course 
within a reliable range), which is considered in our last section before the summary.

\section{Experimental study of superlocalization}
\label{sec_exp}
In this section, we describe the measurement setup and the experimental determination 
of the power absorption as a function of the static (DC) magnetic field for two different 
orientations of the AC and DC fields.

\subsection{Materials and Measurement Method}
\begin{figure}[ht] 
\begin{center} 
\includegraphics[width=5.8cm]{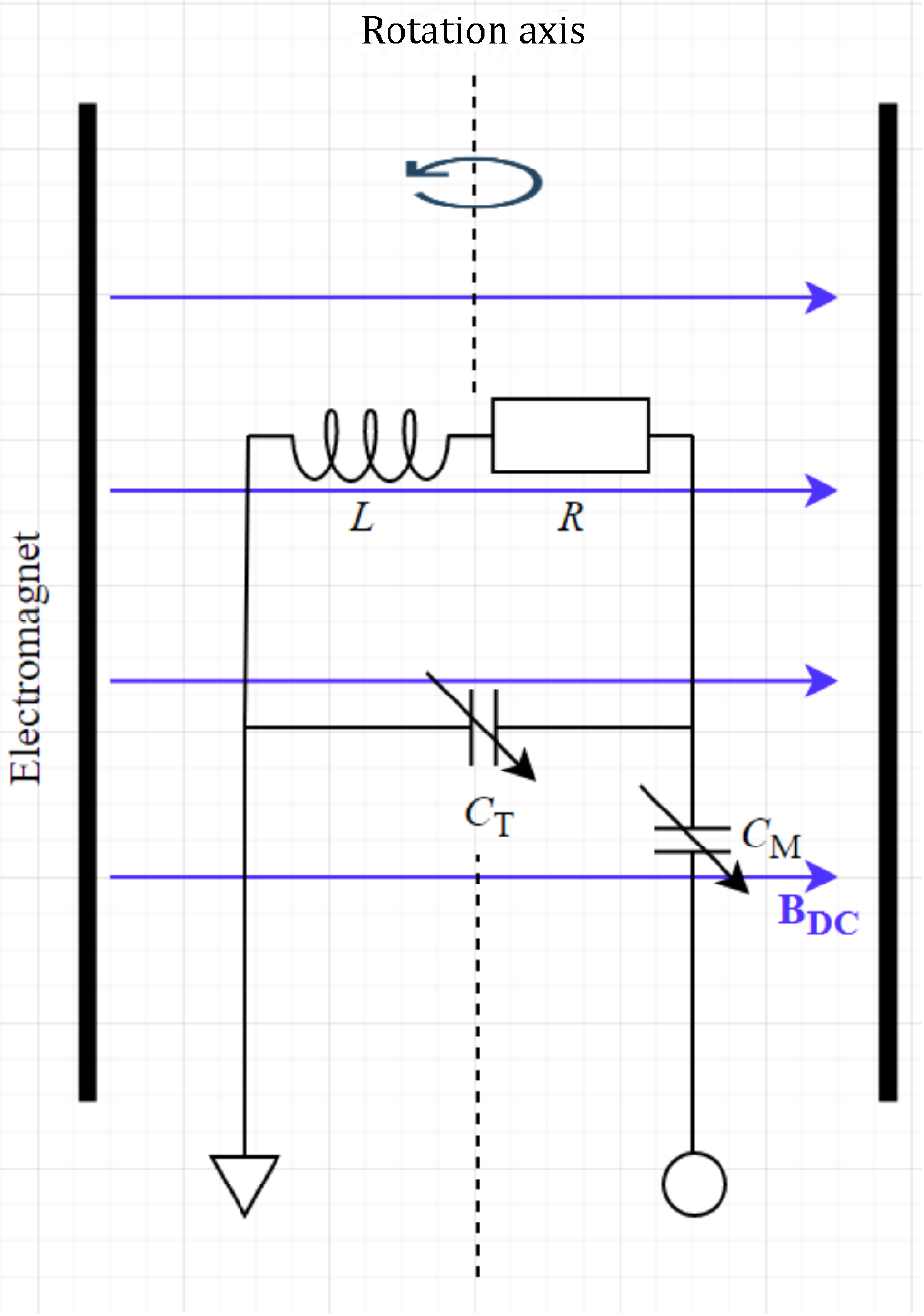}
\caption{(Color online.) The setup of the measuring circuit. The sample is placed inside the coil ($L$) 
and the 2 trimmer capacitors (tuning, $C_{\rm{T}}$, and matching, $C_{\rm{M}}$) are used for setting the 
$50 \,\, \Omega$ impedance at the resonance frequency. It is placed inside the gap of an 
electromagnet. The axis of the coil can be rotated around a vertical axis, the AC magnetic field 
can thus be perpendicular or parallel with respect to the DC magnetic field (blue arrows). 
\label{fig4}
} 
\end{center}
\end{figure}

We used a FERROTEC EMG 705 commercial nanomaterial sample with 65 mg enclosed in a capillary. 
We employed the resonator-based approach to determine the absorbed power as described in 
Refs. \cite{gresits1,gresits2}. The resonant circuit consists of a solenoid coil (containing the sample) 
and 2 trimmer capacitors, presented in \fig{fig4}. 
%
%
The resonator is inside the gap of an electromagnet that can produce DC magnetic field ($B_{\text{DC}}$) 
up to 0.6 T in a stepwise manner. We measured the frequency-dependent reflection from the resonator and 
the resulting resonance curve yields directly the quality factor ($Q$) as a function of $B_{\text{DC}}$. The 
power absorbed by the sample can be calculated from the quality factor \cite{gresits1} as:
\bea
P_{\mr{abs}} = P_{\mr{in}} (1 - Q/Q_{\mr{sat}}), 
\eea
where $P_{\mr{in}}$ is the input power and $Q_{\mr{sat}}$ is the quality factor when the sample 
magnetization is saturated, thus the loss vanishes. We determined that $Q_{\mr{sat}}$ matches that 
of the empty cavity within experimental accuracy. This method is non-invasive and is much less 
time-consuming as compared to a more conventional calorimetric determination of the absorbed power. 
The calorimetric approach requires the placement of a thermometer inside the sample and relatively 
large power densities to achieve rapid sample heating, while the reflectometry can be measured with 
a power of 1 mW or lower. The readout of $Q$ is also instantaneous (faster than 1 sec) while an accurate 
calorimetric measurement requires consecutive heating and cooling cycles. 

The axis of the coil can be rotated around a vertical axis, therefore the direction of the AC magnetic field 
can be perpendicular or parallel with respect to the DC field. We performed the measurements at 2 different 
power levels, 0 dBm (1 mW) and 6 dBm (4 mW). In the latter case, we placed a 6 dB attenuator in front of 
the detector to avoid its saturation and the same detecting power. We did not observe any dependence of 
the observed effects as a function of the AC field strength. Our operating frequency is 35 MHz due to technical 
reasons that is about 2 orders of magnitude larger than the optimal range for hyperthermia as mentioned 
in Sec. \ref{sec_intro}. We also note that the electromagnet has a small remanent magnetic field 
(between 4 and 6 mT depending on the magnetizing history), which however did not affect our conclusions.

\subsection{Experimental Results}
%
%
\begin{figure}[ht] 
\begin{center} 
\includegraphics[width=7.8cm]{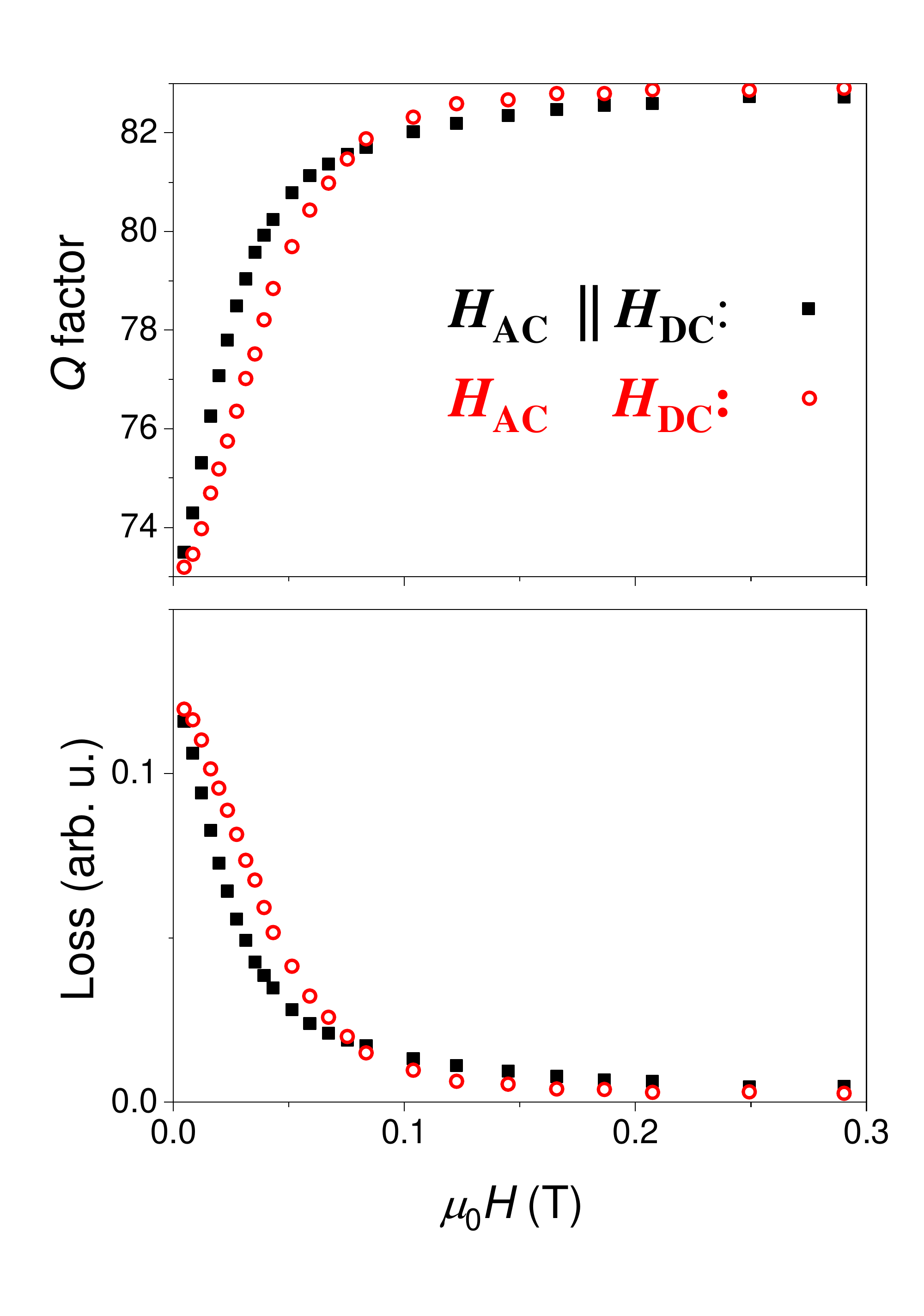}
\caption{(Color online.) The quality factor of the resonator as a function of the DC magnetic field for two 
respective orientations of the DC and AC fields. We observe a saturation of the absorbed power for high 
magnetic fields and also an orientation dependence. The lower panel shows the loss in arbitrary units.
\label{fig5}
} 
\end{center}
\end{figure}

\fig{fig5} shows the change of the quality factor, $Q$ of the resonator as a function of the DC magnetic field. 
This also enabled the determination of the loss as a function of the magnetic field. We observe a gradual saturation 
of $Q$, i.e., a switching-off of the loss above 0.1 T. In addition, the loss data shows a marked anisotropy, i.e., it 
vanishes slower for the geometry when the AC and DC magnetic fields are perpendicular. 

\section{Comparison of theory end experiment}
\label{sec_compare}
\begin{figure}[ht] 
\begin{center} 
\includegraphics[width=7.8cm]{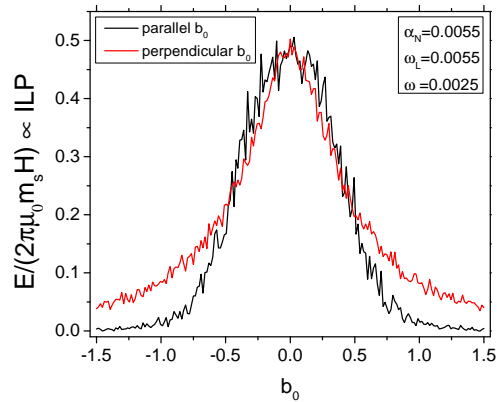}
\caption{(Color online.) Theoretical predictions for $\alpha_N =0.0055, \omega_L = 0.0055, 
\omega=0.0025$ which can be compared directly to the experimental results plotted in \fig{fig5}.
\label{fig6}
} 
\end{center}
\end{figure}

Now we try to put our results into practice and compare the theoretical predictions with the measured data. 
In the theoretical description, we use parameters that approximate the experimental conditions. We choose 
a less powerful oscillating field $H = 1$ kA/m and a higher angular frequency of $\omega = 5 \times 10^7 \, \mr{Hz}$. 
The damping parameter $\alpha$, is not fixed directly by the experimental setup, it is thus a free parameter. 
If one takes $\alpha = 1.0$ (which is close to its usual value and is physically relevant), then the dimensionless 
damping and dimensionless Larmor-frequency becomes equal, $\alpha_N = \omega_L = 0.0055$. In addition, 
the dimensionless angular frequency is given by $\omega = 0.0025$. With these parameters, the theoretical predictions 
for the perpendicular and parallel configurations become similar to each other, as shown in \fig{fig6}.
%
%

We conclude that theoretical predictions in \fig{fig6} are consistent with the experimental results presented in \fig{fig5} 
when the dimensionless damping parameter is displayed to be relatively large, i.e., $\alpha = 1.0$. The overall 
trend in the experimentally observed anisotropic loss can be reproduced with our theoretical description. 
However, the observed superlocalization effect is still much smaller than that predicted for reasonable parameters 
in \fig{fig3}. This is probably related to the relatively large value of $\alpha = 1.0$ and the high experimental frequency.

We believe that a much stronger superlocalization effect on the loss can be observed at lower frequencies and 
for alternative ferrofluids which display a lower damping parameter. This should motivate further experimental 
investigations in this direction.

\section{Summary}
\label{sec_sum}
In summary, we studied the stochastic Landau-Lifshitz-Gilbert (LLG) equations for a particular geometry, i.e.,
when a simultaneously applied AC and DC magnetic fields have varying respective orientations. Our goal was 
to study whether the loss in a hyperthermia-relevant nanomagnetic material can be controlled with an external 
DC magnetic field and its orientation. The theoretical results predict a strong superlocalization effect on the loss. 
This finding is valid for low-frequencies and relatively large field strengths where the linear response
theory cannot be applied. The magnetization vector follows the external magnetic field with almost no phase shift.
Thus, energy loss can only be observed if the external field vanishes. If the AC and DC magnetic fields are parallel 
to each other, the magnitude of the DC field should be smaller than or identical to the AC field to find a zero applied 
field. If the AC and DC magnetic 
fields have a perpendicular orientation, a very small DC field is sufficient to result in a non-vanishing applied field,
so in this case, the superlocalization effect is much stronger than in the parallel case. However, if the frequency
is high and the field strength is small, one expects a linear response, and almost identical superlocalization for the 
two orientations.
This was investigated experimentally using a recently developed accurate measurement method of the loss. 
The experimental data reproduced the theoretically predicted trend for the anisotropy of the superlocalization, 
however with a different magnitude. 
The slight difference between the theoretical calculations and the experimental results can be either due to 
i) the MNPs are assumed to be isotropic and spherical and the slight anisotropy, not treated here, might play a role. 
ii) The Brown relaxation (mechanical rotation of the particle as a whole) was neglected, but this should only yield a 
lower order correction; iii) The material studied in the experiments might have some multidomain particles, as well as 
clustered flakes which cannot be treated easily and beyond the scope of the present investigation;  
(iv) The parameters used in the theoretical predictions of \fig{fig6} and in the experimental results of \fig{fig5} are not 
identical but closer to each other, than those used in \fig{fig3}, so, one expect a better agreement between \fig{fig6} 
and \fig{fig5} which is indeed the case. Nonetheless, the presented theoretical results agree well with previous 
literature at lower frequencies. This can be probably assigned to the large AC frequency in the experiment 
and a large damping parameter.

Regarding future works let us discuss the validity of our findings. We argued that the polarisation effect 
is related to the breakdown of the linear response of the system which happens for relatively large field strength 
and small frequencies. If the field strength is much smaller than the value used by us, linear response could work 
even for low frequencies valid for hyperthermia. Thus, our findings, i.e., the polarised superlocalization may not hold 
in that case. However, in order to maximize the heat transfer, the field strength and the frequency (and their product) 
have to be chosen as high as possible. Therefore, our finding is valid for those cases which are more useful for 
hyperthermia. The detailed study of polarised superlocalization for moderate or small  field strengths are reserved 
for future works.

\section*{Acknowledgement}
The CNR/MTA Italy-Hungary 2019-2021 Joint Project  "Strongly interacting systems in confined geometries" 
and the COST CA17115 - "European network for advancing electromagnetic hyperthermic medical technologies"
are gratefully acknowledged. This work was also supported by the Hungarian National Research, Development 
and Innovation Office (NKFIH) Grant No. K137852, and the Quantum Information National Laboratory sponsored 
by the Ministry of Innovation and Technology via the NKFIH."


\begin{thebibliography}{99}

\bibitem{pankhurst}
Q. A. Pankhurst, J. Connolly, S. K. Jones, J. Dobson, J. Phys. D: Appl. Phys. {\bf 36}, R167 (2003).

\bibitem{ortega_d}
D. Ortega, Q. A. Pankhurst, Nanoscience {\bf 1}, e88 (2003).

\bibitem{pankhurst_progress}
Q. A. Pankhurst,  N. T. K. Thanh, S. K. Jones, J. Dobson, J. Phys. D: Appl. Phys. {\bf 42}, 224001 (2003).

\bibitem{clinapp_1}
P. Wust, U. Gneveckow, M. Johannsen, {\em et al.}, Int. J. Hyperthermia {\bf 22}, 673 (2006).

\bibitem{clinapp_2}
B. Thiesen, A. Jordan,  Int. J. Hyperthermia {\bf 24}, 467 (2008).

\bibitem{perigo_e_a}
E. A. Perigo, G. Hemery, O. Sandre, D. Ortega, E. Garaio, F. Plazaola, F. J. Teran,  Appl. Phys. Rev. {\bf 2}, 041302 (2015).

\bibitem{cabrera_d}
D. Cabrera, J. Camarero, D. Ortega, F. J. Teran, J. Nanopart. Res. {\bf 17}, 121 (2015).

\bibitem{review_mnp}
R. Dhavalikar, A. C. Bohorquez, C. Rinaldi, Micro and Nano Technologies, Pages 265-286 (2019).

\bibitem{mpi_test}
Z. W. Tay, P. Chandrasekharan, A. Chiu-Lam, D. W. Hensley, R. Dhavalikar,  X. Y. Zhou, E. Y. Yu, P. W. Goodwill, B. Zheng, C. Rinaldi, and  S. M. Conolly, 
ACS Nano {\bf 12}, 3699 (2018).

\bibitem{adv_func_mat}
A. Chiu-Lam and C. Rinaldi, Advanced Functional Materials {\bf 26}, 3933 (2016). 

\bibitem{non_local_heat}
C. Kut,  Y. Zhang,  M. Hedayati,  H. Zhou, C. Cornejo,  D. Bordelon,  J. Mihalic,  M. Wabler,  E. Burghardt,  C. Gruettner,  A. Geyh, C. Brayton, T. L. Deweese,  and R. Ivkov, 
Nanomedicine (Lond) {\bf 7}, 1697 (2012).

\bibitem{focused_hyperthermia}
T. O. Tasci, I. Vargel, A. Arat, E. Guzel, P. Korkusuz, E. Atalar, Med. Phys. {\bf 36}, 1906 (2009);
S. L. Ho, L. Jian, W. Gong, W.N. Fu, IEEE Trans. Magn. {\bf 48}, 3262 (2012);
S .L. Ho, S. Niu, W. N. Fu, IEEE Trans. Magn. {\bf 48}, 3254 (2012);
M. Ma, Y. Zhang, X. Shen, J. Xie, Y. Li, N. Gu,  Nano Res. {\bf 8}, 600 (2015).

\bibitem{recent_focused_hyperthermia}
D.W. Hensley, Z.W. Tay, R. Dhavalikar, B. Zheng, P. Goodwill, C. Rinaldi, S. Conolly, Phys. Med. Biol. {\bf 62}, 3483 (2017).

\bibitem{stat_along_increased}
 K. Murase, H. Takata, Y. Takeuchi, S. Saito, Phys. Med. {\bf 29}, 624 (2013).
 
\bibitem{MRSh_theory_focused}
R. Dhavalikar, C. Rinaldi, J. Magn. Magn. Mater. {\bf 419}, 267 (2016).

\bibitem{superlocal_thermal}
Zs. Isz\'aly, I. G. M\'ari\'an, I. A. Szab\'o, A. Trombettoni, and I. N\'andori, J. Magn. Magn. Mater. {\bf 541} 168528 (2022).

\bibitem{LLG}
L. Landau and E. Lifshitz, Phys. Z. Sowjetunion {\bf 8}, 153 (1935);
T. L. Gilbert, Phys. Rev. {\bf 100} (1955) 1243.

\bibitem{sLLG}
W. F. Brown, Phys. Rev. {\bf 130}, 1677 (1963);
W. F. Brown, IEEE Trans. Magn. Magn-15 (1979).

\bibitem{usov_hysteresis}
N. A. Usov, J. Appl. Phys. {\bf 107}, 123909 (2010).

\bibitem{usov_claster}
N. A. Usov, O. N. Serebryakova, and V. P. Tarasov, Nanoscale Res. Lett. {\bf 12}, 489 (2017). 

\bibitem{lyutyy_general}
T. V. Lyutyy, O. M. Hryshko, A. A. Kovner,  J. Magn. Magn. Mater. {\bf 446}, 87-94 (2018).

\bibitem{neel_brown} 
P. T. Phong, L. H. Nguyen, I. J. Lee, N. X. Phuc,  J. Electron. Mater. {\bf 46}, 2393-2405 (2017).

\bibitem{viscous_rotating}
N. A. Usov, R. A. Rytov, V. A. Bautin,  Beilstein J. Nanotechnol. {\bf 10}, 2294 (2019).

\bibitem{hergt_dutz}
R. Hergt, W. Andra, C. G. d'Ambly, I. Hilger, W. A. Kaiser, U. Richter, H.-G. Schmidt, IEEE Trans. Magn. {\bf 34}, 3745 (1998);
R. Hergt, S. Dutz, M. Zeisberger, Nanotechnology {\bf 21}, 015706 (2010);
R. Hergt, S. Dutz,  J. Magn. Magn. Mater. {\bf 311}, 187 (2007).

\bibitem{magforce}
U. Gneveckow, A. Jordan, R. Scholz, V. Bruss, N. Waldofner, J. Ricke, A. Feussner, B. Hildebrandt, B. Rau, P. Wust, Med. Phys. {\bf 31} 1444 (2004);
B. Thiesen, A. Jordan, Int. J. Hyperth. {\bf 24} 467 (2008).

\bibitem{high_amplitude}
D. Bordelon, C. Cornejo, C. Gr\"uttner, F. Westphal, T. L. DeWeese, R. Ivkov, J. Appl. Phys. {\bf 109}, 124904 (2011).

\bibitem{Fannin}
P. C. Fannin, I. Malaescue, C. N. Marin,  J. Magn. Magn. Mater. {\bf 289}, 162 (2005).

\bibitem{thermal_exp}
W. Wernsdorfer, {\it et. al}, Phys. Rev. Lett. {\bf 78}, 1791 (1997).

\bibitem{thermal_summary}
W.T. Coffey, Y.P. Kalmykov, J., App. Phys. {\bf 112}, 121301 (2012).

\bibitem{path_int_sllg}
Camille Aron,  {\em et al.}, J. Stat. Mech. {\bf 2014}, 09008 (2014).

\bibitem{Lyutyy_energy}
T. V. Lyutyy, S. I. Denisov, A. Yu. Peletskyi and C. Binns, Phys. Rev. B. {\bf 91}, 054425 (2015). 

\bibitem{Giordano}
S. Giordano, Y. Dusch, N. Tiercelin, P. Pernod, V. Preobrazhensky, Eur. Phys. J. B {\bf 86}, 249 (2013). 

\bibitem{gresits1}
I. Gresits, Gy. Thur\'oczy, O. S\'agi, B. Gy\"ure-Garami, B. G. M\'arkus and F. Simon, Sci. Rep. {\bf 8}, 12667 (2018).

\bibitem{gresits2}
I. Gresits, Gy. Thur\'oczy, O. S\'agi, I. Homolya, G. Bagam\'ery, D. Gaj\'ari, M. Babos, P. Major, B. G. M\'arkus and F. Simon,  J. Phys. D: Appl. Phys. {\bf 52} 375401 (2019).

\bibitem{ferguson}
R. M. Ferguson, {\it et al.} IEEE Trans Magn. {\bf 49}, 3441 (2013).

\bibitem{gresits3}
I. Gresits, Gy. Thur\'oczy, O. S\'agi, S. Kollarics, G. Cs\H{o}sz, B.G. M\'arkus, N.M. Nemes, M. Garc\'ia Hern\'andez and F. Simon, J. Magn. Magn. Mater. {\bf 526}, 167682 (2021).

\bibitem{ota}
Satoshi Ota, and Yasushi Takemura, J. Phys. Chem. C {\bf 123}, 28859 (2019).

\end{thebibliography}
\end{document}